\documentclass[aps,prd,eqsecnum,showpacs,amsmath,nofootinbib,superscriptaddress,twocolumn,floats,preprintnumbers]{revtex4}

\usepackage[dvips]{color,graphicx}
\usepackage{amsfonts,amssymb,theorem,mathrsfs,times}
\definecolor{red  }{rgb}{1,0,0}
\definecolor{blue }{rgb}{0,0,1}
\definecolor{green}{rgb}{0,1,0}

\begin{document}
\newcommand{\M}{\mathbb{M}}
\newcommand{\R}{\mathbb{R}}
\newcommand{\HI}{\mathbb{H}}
\newcommand{\C}{\mathbb{C}}
\newcommand{\TI}{\mathbb{T}}
\newcommand{\E}{\mathbb{E}}
\def\etal{{\it et al}.} \def\e{{\rm e}} \def\de{\delta}
\def\dd{{\rm d}} \def\ds{\dd s} \def\ep{\epsilon} \def\de{\delta}
\def\goesas{\mathop{\sim}\limits} \def\al{\alpha} \def\vph{\varphi}
\def\Z#1{_{\lower2pt\hbox{$\scriptstyle#1$}}}

\newcommand{\be}{\begin{equation}}
\newcommand{\ee}{\end{equation}}
\newcommand{\bea}{\begin{eqnarray}}
\newcommand{\eea}{\end{eqnarray}}
\newcommand{\nn}{\nonumber}

\def\IR{{\hbox{{\rm I}\kern-.2em\hbox{\rm R}}}}


\renewcommand{\thefootnote}{\fnsymbol{footnote}}
\long\def\@makefntext#1{\parindent 0cm\noindent \hbox to
1em{\hss$^{\@thefnmark}$}#1}
\def\Z#1{_{\lower2pt\hbox{$\scriptstyle#1$}}}


\title{De Sitter brane-world, localization of gravity, and the cosmological constant}

\preprint{UOC-TP 014/10}

\author{Ishwaree P. Neupane}
\affiliation{Department of Physics and Astronomy, University of
Canterbury, Private Bag 4800, Christchurch 8041, New Zealand}

\begin{abstract}

Cosmological models with a de Sitter 3-brane embedded in a
5-dimensional de Sitter spacetime (dS$_5$) give rise to a finite
4D Planck mass similar to that in Randall-Sundrum (RS) brane-world
models in anti--de Sitter 5-dimensional spacetime (AdS$_5$). Yet,
there arise a few important differences as compared to the results
with a flat 3-brane or 4D Minkowski spacetime. For example, the
mass reduction formula (MRF) $M_{\rm Pl}^2=M_{(5)}^3 \ell_{\rm
AdS}$ as well as the relationship $M_{\rm Pl}^2= M_{{\rm
Pl}(4+n)}^{n+2} L^{n}$ (with $L$ being the average size or the
radius of the $n$ extra dimensions) expected in models of
product-space (or Kaluza-Klein) compactifications get modified in
cosmological backgrounds. In an expanding universe, a physically
relevant MRF encodes information upon the 4-dimensional Hubble
expansion parameter, in addition to the length and mass parameters
$L$, $M_{\rm Pl}$ and $M_{{\rm Pl}(4+n)}$. If a bulk cosmological
constant is present in the solution, then the reduction formula is
further modified. With these new insights, we show that the
localization of a massless 4D graviton as well as the mass
hierarchy between $M_{\rm Pl}$ and $M_{{\rm Pl} (4+n)}$ can be
explained in cosmological brane-world models. A notable advantage
of having a 5D de Sitter bulk is that in this case the zero-mass
wave function is normalizable, which is not necessarily the case
if the bulk spacetime is anti de Sitter. In spacetime dimensions
$D\ge 7$, however, the bulk cosmological constant $\Lambda\Z{b}$
can take either sign ($\Lambda\Z{b} <0$, $=0$, or $>0$). The $D=6$
case is rather inconclusive, in which case $\Lambda_b$ may be
introduced together with 2-form gauge field (or flux). We obtain
some interesting classical gravity solutions that compactify
higher-dimensional spacetime to produce a Robertson-Walker
universe with de Sitter-type expansion plus one extra noncompact
direction. We also show that such models can admit both an
effective 4-dimensional Newton constant that remains finite and a
normalizable zero-mode graviton wave function.

\end{abstract}

\pacs{11.25.Mj, 04.65.+e, 11.25.Yb, 98.80.Cq,}

\maketitle


\section{Introduction}

The Universe is endowed with a number of cosmological mysteries, but
the one that most vexes physicists is the smallness of the observed
vacuum energy density in the present Universe and its effects on
an accelerated expansion of the Universe at a late
epoch~\cite{supernovae}. This cosmological enigma has so far
defied an elegant and forthright explanation.

Brane-worlds are promising theories with extra spatial dimensions
in which ordinary matter is localized on a (3+1)-dimensional
subspace~\cite{Domain-wall}. To this end, the Randall-Sundrum (RS)
models in 5 dimensions~\cite{RS1,RS2} could be viewed as the
simplest brane-world configurations with 1 extra dimension of
space. The RS models and their generalizations in higher spacetime
dimensions are known to have interesting consequences for
gravitational physics~\cite{Shiromizu:1999wj,
Garriga99,Chamblin99,Neupane:2001} and cosmology, see,
e.g.,~\cite{Csaki99,Cline99,Binetruy:1999,Mukohyama99c}; we refer
to~\cite{Maartens:2010} for review and further references.

In the simplest Randall-Sundrum brane-world models, one has a flat
3-brane (or a 4D Minkowski spacetime) embedded in a
5-dimensional anti-de Sitter spacetime, known as an AdS$_5$
bulk. In this simple setting there exist a massless graviton (or
zero-mode) and massive gravitons (or Kaluza-Klein modes) of metric
tensor fluctuations. The massless graviton mode reproduces the
standard Newtonian gravity on the 3-brane, while the Kaluza-Klein
modes, which arise as the effect of graviton fluctuations in extra
dimension(s), give corrections to the Newton's force
law~\cite{Garriga99}. The 5D bulk geometry is extremely warped in
these models, as is reflected from a typical size of the 5D
curvature radius, $\ell_{\rm AdS} < 0.1~{\rm mm}$. Consequently
the Newtonian gravity is recovered at distances larger than ${\cal
O}(0.1~{\rm mm})$.

The requirement of an AdS$_5$ bulk spacetime in the original RS
brane-world models may {\it not} be something that is totally
unexpected since certain versions of 10D string theory,
particularly, type IIB string theory, are known to contain AdS$_5$
space as a sub-background space of the full spacetime, which is
${\rm AdS}_5 \times S^5$, and string theory itself is viewed as
the most promising candidate for the unified theory of everything.
However, the original RS models also predict a zero cosmological
constant on the brane or 4D spacetime. This result is not
supported by cosmological observations, which favor a positive
cosmological constant-like term in 4 dimensions.

To construct a natural theory of brane-world, we shall replace the
flat 3-brane of the original RS setup by a dynamical brane or a
physical $3+1$-dimensional hypersurface with a nonzero Hubble
expansion parameter, for instance, by a
Friedmann-Lama\^itre-Robertson-Walker metric. With such a simple
modification of the original RS brane-world model, the zero-mode
graviton fluctuation is not guaranteed to be localized on the
brane, if the 5D bulk spacetime is anti--de Sitter. However, if the
5D bulk spacetime is de Sitter or positively curved, then there
always exists a normalizable zero-mode graviton localized on a de
Sitter brane. In such theories the smallness of the 4D
cosmological constant term can be related to an infinitely large
extension of the fifth dimension.

There is another motivation for considering a positively curved 5D
background spacetime. When we consider compactifications of
string/M-theory or classical supergravity theories with more than
one extra dimension, then in a cosmological setting, and under
the dimensional reduction from D dimensions to 5, we generally
find that the 5D spacetime is de Sitter, if we also insist on the
existence of a 4-dimensional de Sitter solution. Even though
AdS$_5$ is well--motivated from some aspects of type IIB
supergravity, for its role in the AdS/CFT correspondence, it is
difficult to realize an AdS$_5$ background, while at the same time
we also obtain a dS$_4$ solution (or an inflating FRW universe in
4 dimensions) by solving the full D-dimensional Einstein
equations.

In this paper we show how brane-world models with a positively
curved bulk spacetime (dS$_5$) can generate a 4-dimensional
cosmological constant in the gravity sector of the effective 4D
theory with a finite 4D Newtons constant and also help explain the
localization of a normalizable zero-mode graviton in 4
dimensions. We also present some new insights on localization of
gravity on a de Sitter brane embedded in a higher-dimensional bulk
spacetime.

In Sec. II, we pay particular attention in the study of classical
gravity solutions in a 5D de Sitter bulk spacetime (dS$_5$) and
derivation of the associated mass reduction formulae. The idea of
replacing an AdS$_5$ bulk spacetime by a dS$_5$ spacetime is not
totally new; the behavior of classical gravity solutions on a
dS$_4$ brane embedded in dS$_5$ was discussed before, e.g.,
in~\cite{Brevik-etal,Kehagias02}. Certain aspects of linearized
gravity in a dS$_5$ bulk spacetime were also discussed in these
papers, mainly, the limits between which the value of the bulk
cosmological constant has to lie in order to localize the graviton
on a de Sitter 3-brane~\cite{Brevik-etal} and the correction to
the static potential at short distances due to the massive KK
modes living in a 5D bulk spacetime~\cite{Kehagias02}. Part of our
analysis in Sec. III finds a similarity with earlier works in
the subject, but the details and some of our conclusions are
different. For instance, for an embedding of dS$_4$ brane into
AdS$_5$ bulk, we find that the massless graviton wave function is
non-normalizable (despite being a bound state solution) once the
$Z_2$ symmetry is relaxed. This result is consistent with our
observation in Sec. II that the 4D Newton's constant is not
finite in the absence of $Z_2$ symmetry. A more detailed
comparison between past and present results will be made in the
context of the analysis below.

Different from the approaches in~\cite{Brevik-etal,Kehagias02},
our analysis will be based on conformal coordinates (rather than
on Gaussian normal coordinates) in terms of which the discussions
of classical solutions and localization of gravity become simpler
and more precise. For the sake of completeness, we shall also
analyze a set of linearized bulk equations in a dS$_5$ spacetime
(the case of AdS$_5$ bulk was considered before, for example,
in~\cite{Sasaki2000a,Koyama:04a}. Furthermore, we provide new
insights on the nature of mass gap in higher dimensions, by
allowing more than one extra dimensions. Our focus in this paper
will remain on the discussion of mass reduction formulas and also
on the behavior of classical gravity solutions in higher
dimensions. These were not discussed before in the literature,
including Refs.~\cite{Brevik-etal,Kehagias02}, at least,
at the level of clarity and details as have been done in this
paper. As a canonical example in 5 dimensions, we also show
that the effective 4D Newton's constant can be finite for gravity
theories coupled to a bulk scalar field, despite having a
noncompact direction.

In Sec. IV we consider classical gravity solutions in higher
dimensions. In dimensions $D\ge 7$, we find that the bulk
cosmological term $\Lambda\Z{b}$ can take either sign, though
a negative $\Lambda\Z{b}$ may be preferred over a positive
$\Lambda\Z{b}$ for regularity of the metric. We will also make
some general remarks about mass reduction formulas (MRFs). In the
standard Kaluza-Klein theories, the MRF is given by $M_{\rm Pl}^2=
M_{{\rm Pl}(4+n)}^{n+2} L^{n}$, which relates the 4-dimensional
effective Planck mass $M_{\rm Pl}$ with the $(4+n)$-dimensional
Planck mass $M_{{\rm Pl} (4+n)}$ (with $L$ being the average size
of the $n$ extra dimensions). This result, also known as Gauss
formula, gets naturally modified in the presence of a bulk
cosmological term and also due a nonzero 4-dimensional Hubble
expansion parameter. Finally, in Sec. V we will study a
particular class of gravity solutions in the presence of a bulk
scalar field. Concluding remarks are given in Sec. V.

The important contributions of this paper include (i)
generalizations of the 5D results in higher dimensions, (ii) clear
and more precise interpretations of the relationships between the
4D effective Planck mass and the D-dimensional fundamental mass
scales, which take into account the effect of cosmic expansion or
the Hubble expansion parameter plus the bulk cosmological constant
term, and (iii) several new aspects of gravity localization both
in 5D and higher dimensions.


\section{De Sitter brane-worlds}

5-dimensional de Sitter brane-worlds characterized by a single
extra dimension where the bulk spacetime is positively curved
(instead of being flat or negatively curved) are among some highly
plausible approaches to explaining the smallness of the observed
cosmological vacuum energy density and localization gravity.

The basic idea behind the existence of a 4-dimensional de
Sitter space solution (${\rm dS}_4$) supported by warping of extra
spaces can be illustrated by considering a curved 5-dimensional
``warped metric",
\begin{equation}\label{5d-main-ac}
ds_{5}^2 = e^{2 A(\phi)} \left( ds_4^2 + \rho^2 d\phi^2\right),
\end{equation}
where $\rho$ is a free parameter with dimension of length and
$e^{2A(\phi)}$ is the warp factor as a function of $\phi$. We look
for solutions for which the 4-dimensional line--element takes
the standard Friedmann-Lama\^itre-Robertson-Walker (FLRW) form
\begin{eqnarray}\label{FRW}
ds_4^2 &\equiv&  {g}_{\mu\nu} dx^\mu dx^\nu\nonumber \\
&=&  - dt^2+ a^2(t)\left[\frac{ dr^2}{1-\kappa r^2}+ r^2
d\Omega_2^2\right],
\end{eqnarray}
where $\kappa$ is the 3D curvature constant with the dimension of
inverse length squared. The 5D background Riemann tensor satisfies
\begin{equation}
 {}^{(5)} R_{ABCD} = \frac{\Lambda_5}{6} \left( {}^{(5)} g\Z{AC}
{}^{(5)} g\Z{BD}-{}^{(5)} g\Z{AD}{}^{(5)} g\Z{BC}\right).
\end{equation}
The 5D Einstein-Hilbert action takes the form
\begin{equation}\label{5d-gravi}
S_{\rm grav}  = M_{(5)}^3 \int d^5{x} \sqrt{-g}
\left(R-2\Lambda_5\right),
\end{equation}
where $M_{(5)}$ is the 5D Planck mass and $\Lambda_5\equiv
6/\ell^2$ and $\ell$ is the radius of curvature of the 5D bulk
spacetime.

The gravitational action (\ref{5d-gravi}) may be supplemented with
the following 3-brane action
\begin{equation}
S_{\rm brane} = \int_{\partial {\cal M}} \sqrt{-g_{b}}\,(-\tau),
\end{equation}
where $\tau$ denotes the brane tension. The 5D Einstein field
equations are given by
\begin{equation}\label{5DEinstein}
G_{AB}= - \frac{\tau}{2} \frac{\sqrt{-g_b}}{\sqrt{-g}}
g_{\mu\nu}^b \delta_A^\mu \delta_B^\nu \delta(\phi-\phi_0)
-\Lambda_5 g_{AB}.
\end{equation}
The 3 independent equations of motion are
\begin{eqnarray}
\frac{6{A'}^2}{\rho^2} &=& 6\left(\frac{\dot{a}^2}{a^2}+
\frac{\kappa}{a^2}\right) -\Lambda_5\, e^{2A},\\
\frac{6A''}{\rho^2} &=& - \Lambda_5\, e^{2A} -\frac{\tau}{\rho
M_{(5)}^3}\, \delta(\phi-\phi_0)
e^{A},\\
\frac{\ddot{a}}{a} &=& \frac{\dot{a}^2}{a^2} + \frac{\kappa}{a^2}.
\end{eqnarray}
Here we are interested in studying a theory without the orbifold
boundary condition, that is a theory with infinite extent in both
the positive and negative $\phi$ direction. We allow both even and
odd functions of $\phi$ rather than the restriction to purely even
functions demanded by the orbifold conditions in RS brane-world
models.

\subsection{A spatially flat universe}

First, we take $\kappa=0$ (spatially flat universe): The 5D
Einstein equations are solved with the scale factor
\begin{equation}
a(t)= a_0 \, e^{H t}
\end{equation}
and the warp factor
\begin{equation}\label{sol-bg1}
A(\phi)= \ln (2 \ell_0 H) - \ln \left(\exp(\rho H \phi)+
\frac{\ell_0^2}{\ell^2}\exp(-\rho H \phi)\right), \end{equation}
where $\ell_0$ and $H$ are two integration constants. The standard
results in AdS$_5$ space, see, for
example~\cite{Sasaki2000a,Ish09a,Ish09b}, are obtained by
replacing $\ell^2$ with $-\ell_{\rm AdS}^2$ or $6/\Lambda_5$.

From the explicit solution given above, we derive
\begin{eqnarray}\label{4d-eff-act}
S_{\rm eff}^{(D=4)} = M_{\rm Pl}^2 \int d^4{x} \sqrt{-g_4}  \,
{R}_4 - \int d^4{x} \sqrt{-{g}_4}\, K,
\end{eqnarray}
where
\begin{eqnarray}\label{4d-planck}
M_{\rm Pl}^2 &=& 8\rho M_{(5)}^3 \ell_0^3 H^3
\int_{-\infty}^{\infty} \left( \e^{\,\rho H\phi} +
\frac{\ell_0^2}{\ell^2} \, \e^{- \rho H \phi}\right)^{-3} d\phi
\nonumber\\
 &= & 8 M_5^3 \ell_0^3 H^2
\frac{\tan^{-1}(\ell_0/\ell)
+\cot^{-1}(\ell_0/\ell)}{8\ell_0^3/\ell^3}\nonumber
\\ &=& \frac{\pi}{2} M_{(5)}^3
\ell^3 H^2,
\end{eqnarray}
\begin{eqnarray}
K  &\equiv& \frac{M_{(5)}^3}{\rho} \int_\infty^{+\infty} e^{3A}
\left(12 {A^\prime}^2+ 8A^{\prime\prime}
+\frac{12\rho^2}{\ell^2}\, e^{2A}\right)d\phi\nonumber \\
&=& 8 M_{(5)}^3 \ell_0^3 H^4 \int_{-\infty}^{\infty} \frac{
4\left(3\,e^{2\, \varphi} +3\lambda^2 \,e^{-2\, \varphi}
-2\lambda\right)}{ \left( e^{ \varphi}+ \lambda \,e^{- \varphi}
\right)^5}\,d\varphi \nonumber \\
&\equiv & 8 M_{(5)}^3 \ell_0^3 H^4 \int_{-\infty}^{\infty}
\Lambda(\varphi),
\end{eqnarray}
where we defined $\varphi\equiv \rho H \phi$ and $\lambda\equiv
\ell_0^2/\ell^2$. This yields
\begin{equation}
K= 8 M_{(5)}^3 \ell_0^3 H^4 \frac{3\pi}{8\lambda^{3/2}}=6 H^2
M_{\rm Pl}^2.
\end{equation}
A similar result was obtained in~\cite{Aguilar10a}, taking
$\lambda=1 $ and $\rho=1$. We find interest only on smooth
brane-world solutions, so we take $\lambda\equiv
\ell_0^2/\ell^2>0$.

\begin{figure}[!ht]
\centerline{\includegraphics[width=2.4in,height=1.5in]
{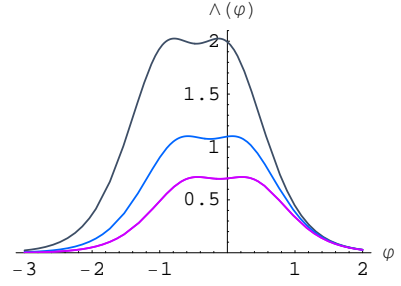}} \caption{ (color online). The plot of the function
$\Lambda(\varphi)$ with $\lambda=0.4, 0.6$, and $0.8$ (from top to
bottom) (black, blue, and pink underline{online}). Like the warp factor
$e^A$, $\Lambda(\varphi)$ is regular, has a peak at $\phi\equiv
\phi_0$ and falls off rapidly away from the brane.} \label{warp}
\end{figure}
The 4-dimensional effective action is given by
\begin{equation}
S_{\rm eff}^{(D=4)}= M_{\rm Pl}^2 \int d^4{x} \sqrt{-g_4}
\left(R_4- \Lambda_{4}\right),
\end{equation}
where the 4D effective cosmological constant
\begin{equation}
\Lambda_{4}= 6 H^2. \end{equation} 
These results are different from those in the simplest RS brane-worlds
at least in 2 aspects. Firstly, the cosmological mass
reduction formula (cMRF)
\begin{equation}\label{5to4Planck}
M_{(5)}^3 =\frac{2 M_{\rm Pl}^2}{\pi \ell^3 H^2}
\end{equation}
is clearly different from that obtained in a static AdS$_5$
brane-world configuration~\cite{RS2}, for which
\begin{equation}
M_{(5)}^3 = \frac{M_{\rm Pl}^2}{\ell_{\rm AdS}}.
\end{equation}
Secondly, and perhaps more importantly, no parameter of the action
would have to be tuned to keep $\Lambda_4$ positive. The
cosmological constant problem of the original RS brane-world
proposal, which is the question of why the background warps in the
appropriate fashion without introducing an effective 4D
cosmological constant, does not arise here. In fact, in a
dynamical spacetime, the vacuum energy on the brane can naturally
warp the bulk spacetime and introduce a nontrivial curvature in
the bulk while maintaining a 4D de Sitter solution.

\subsection{Inclusion of brane action}

In the presence of a brane action, we have to consider the metric
a step function in $\phi$, while computing derivatives of
$A(\phi)$ (with respect to $\phi$). The solution valid for
$-\infty \le \phi \le +\infty $ then implies that
\begin{equation}\label{two-deri}
A^{\prime\prime}+ \frac{4\lambda \rho^2 H^2}{\Phi_+^2} +\frac{\rho
H\Phi_-}{\Phi_+} \left(2\delta (\phi-\phi_0) \right)=0,
\end{equation}
where ${}^\prime\equiv \frac{d}{d\phi}$ and $\Phi_{\pm} \equiv
e^{\rho H\phi } \pm  \lambda \, e^{- \rho H\phi}$. One could think
of a de Sitter brane as the location $\phi=\phi_0$ ($z=z_c$) where
the zero-mode graviton wave function is peaked.

The $\mu\nu$ components of the 5D Einstein equations yield
\begin{eqnarray}
A^{\prime\prime}+ \frac{4\lambda\rho^2 H^2}{\Phi_+^2} +\frac{
2\rho H \ell_0}{3 M_5^3\, \Phi_+} \left(\tau \delta
(\phi-\phi_0) \right)= 0.\label{b-tensions}
\end{eqnarray}
By comparing Eqs.~(\ref{two-deri}) and (\ref{b-tensions}), we get
\begin{equation}\label{brane-ten}
\tau=\frac{3 M_5^3}{\ell_0} \left(e^{\rho
H\phi_0}-\lambda\,e^{-\rho H \phi_0} \right).
\end{equation}
Particularly, in the limit $\lambda \to 0$, the 5D spacetime
becomes spatially flat and gravity is {\it not} localized in this
case. Indeed, $\lambda>0$ is required to keep the warp factor
bounded from the below and above (cf. Figure~\ref{warp}).

For $\lambda>0$, by writing
$$
\phi \equiv \frac{1}{\rho} \left(z + \ln
\frac{\ell_0}{\ell}\right).$$ 
the solution for warp factor, Eq.~(\ref{sol-bg1}), and the brane
tension can be written in standard forms, i.e.
\begin{equation}
e^{A(z)} =\frac{\ell H}{\cosh H z}, \quad \tau=\frac{6
M_{(5)}^3}{\ell}\sinh H z_c,\label{flat-brane-T}
\end{equation}
where $z_c>0$. The scale of warped compactification is
\begin{equation} r_{c} \equiv \rho\, e^{A}=
\frac{2\ell_0\rho H}{\left(e^{\rho H\phi}+\lambda e^{-\rho
H\phi}\right)}=\frac{\rho H \ell}{\cosh {H z}}.
\end{equation}
The $\rho\to \infty$ limit gives rise to a theory with a
semi-infinite extra dimension.

In Sec. IIB of~\cite{Brevik-etal}, the authors presented
5D solutions in terms of Gaussian normal coordinates, which admit
a bulk singularity at $|y|=y_H$, especially, in the $\Lambda_5>0$
case, so there one required a cut-off in the bulk, which may be
seen as the position of the horizon in the bulk. The bulk
singularity at $|y|=y_H$, in terms of Gaussian normal coordinates,
is not a physical singularity. Note that, in terms of the
conformal coordinate $z$, the classical solution presented above
is regular everywhere, particularly, when $\Lambda_5>0$. In the
notation of~\cite{Brevik-etal}, $\lambda\equiv  H^2$, while in our
notation $\lambda\equiv \ell_0^2/\ell^2 = \Lambda_5 \ell_0^2/6$.

\subsection{Nonflat universe}

In a spatially nonflat universe ($\kappa\ne 0$), the 5D Einstein
equations are explicitly solved when
\begin{equation}
a(t)=\frac{c\Z{0}^2+\kappa \rho^2}{2 c\Z{0}}\cosh \left(\frac{
t}{\rho}\right) + \frac{c\Z{0}^2-\kappa \rho^2}{2 c\Z{0}}\sinh
\left(\frac{ t}{\rho}\right)
\end{equation}
and
\begin{equation}
A(\phi)= \ln \left(\frac{2\ell_0}{\rho}\right) - \ln \left(
\exp(\phi)+ \frac{\ell_0^2}{\ell^2} \exp (-\phi) \right),
\end{equation} The Hubble-like parameter $H$
(which appeared in the $\kappa=0$ case above) is no more arbitrary but
it is fixed in terms of the length parameter $\rho$, i.e., $H\to
1/\rho$. The 4D effective action still takes the form of
(\ref{4d-eff-act}), but now
\begin{eqnarray}\label{Planck-sol2}
M_{\rm Pl}^2 =\frac{\pi}{2} \frac{M_{(5)}^3 \ell^3}{\rho^2},
\qquad  K=\frac{3\pi M_{(5)}^3\ell^3}{\rho^4}.
\end{eqnarray}
Note that, unlike in the simplest RS brane-world models, we do not
require the ${Z}_2$ symmetry in order to get a finite 4D Planck
mass, as long as $\lambda>0$ or $\ell_0^2/\ell^2>0$.

In the $\kappa\ne 0$ case, Eq.~(\ref{brane-ten}) is modified
as
\begin{equation}
\tau=\frac{3 M_5^3}{\ell_0}
\left(e^{\phi_0}-\lambda\,e^{-\phi_0} \right).
\end{equation}
The scale of warped compactification is now
\begin{equation} r_{c} \equiv \rho\, e^{A}=
\frac{2\ell_0}{\left(e^{\phi}+\lambda e^{-\phi}\right)}\equiv
\frac{2\ell}{\cosh y},
\end{equation}
where $y\equiv \phi-\ln(\ell_0/\ell)$. This is exponentially
suppressed as $y\to \pm \infty$. Clearly, there is no problem with
taking the $\rho\to \infty$ limit of the background solution given
above.

\medskip

\section{Linearized gravity in 5D}

Brane-world models with one or more noncompact extra spaces are
known to require the trapping of gravitational degrees of freedom
on the brane~\cite{RS2,Neupane:2001}. To determine whether the
spectrum of linearized tensor fluctuations $\delta {}^{(5)}
g_{AB}$ is consistent with 4D experimental gravity, we shall
consider the perturbations around the background solution given
above.

The $\kappa\ne 0$ solutions are slightly more restrictive than the
$\kappa=0$ solutions. So, henceforth, we focus our discussions on
the $\kappa=0$ case, for which $\rho$ is arbitrary. The
perturbations of the 5D metric $\delta^{(5)} g_{AB}\equiv h_{AB}$
may be written as
\begin{widetext}
\begin{equation}
  \renewcommand{\arraystretch}{1.5}
  \delta \:{}^{(5)\!}g_{AB} = \left[
    \begin{array}{@{\quad}c@{\quad}c@{\quad}|@{\quad}c@{\quad}}
      -2 e^{2A} \psi &
      e^{2A} a^2 (\partial_i{\cal B}-S_i) &
      e^{A}\xi
      \\
      e^{2A} a^2 (\partial_j{\cal B}-S_j) &
      e^{2A} a^2 \left\{2{\cal R} \delta_{ij} +
      2\partial_i \partial_j{\cal C} +
      2\partial_{(i} V_{j)}+ h_{ij} \right\} &
      e^{2A} a^2 (\partial_i\beta-\chi_i) \\ [0.5 em]
      \hline
      e^{A} \xi &
      e^{2A} a^2 (\partial_j\beta-\chi_j) &
      2 \rho^2 e^{2A} \zeta
    \end{array}
  \right],
  \label{pertmetric}
\end{equation}
where $\psi, {\cal R}, {\cal C}, \zeta, \beta, \xi $ are metric
scalars,  while $S_i, V_i, \chi_i$ are transverse 3D vector
fields, and $h_{ij}$ represent transverse-traceless tensor modes.
Here we focus on the analysis of tensor modes (we refer
to~\cite{Koyama:04a} for the analysis of gauge-invariant scalar
and vector perturbations of maximally symmetric spacetimes), see
also~\cite{Langlois00a}.
The transverse-traceless tensor modes $h_{ij}\equiv \delta
g_{ij}=\delta_i^\mu \delta_j^\nu h_{\mu\nu}(x,\phi)$ satisfy the
following wave equation
\begin{eqnarray}
&& e^{-2A} \left( \frac{1}{\rho^2} \frac{\partial}{\partial\phi^2}
+\frac{3A'}{\rho^2}  \partial_\phi - \partial_t^2 - 3
\frac{\dot{a}}{a}
\partial_t + \frac{\vec{\nabla}^2}{a^2}\right)h_{ij} + \frac{\tau}{2 M_{(5)}^3} \frac{e^{-A}}{\rho}
\delta(\phi-\phi_0)h_{ij}=0.
\end{eqnarray}
\end{widetext}
The last term above has arisen from the first term on the right
hand side in Eq.~(\ref{5DEinstein}). By separating the variables
as
\begin{equation}
h_{ij} ( x^\mu, \phi) \equiv \sum \alpha_m(t) u_m(\phi)\,e^{i
k\cdot x}\,\hat{e}_{ij},
\end{equation}
where $e_{ij} (x^i)$ is a transverse, tracefree harmonics on the
spatially flat 3-space, $\vec{\nabla}^2 \hat{e}_{ij} = - k^2
\hat{e}_{ij}$, we get
\begin{subequations}
\begin{align}
\ddot{\alpha}_m + 3 \frac{\dot{a}}{a} \dot{\alpha}_m + \left(
  \frac{k^2}{a^2}+m^2 \right) \alpha_m=0,\label{TT-scalar}\\
\left(\frac{1}{\rho^2}\frac{d^2}{d\phi^2} +
\frac{3A^\prime}{\rho^2} \frac{d}{d\phi}+\frac{3 H \Phi_-}
{\rho\Phi_+}\,\delta(\phi-\phi_0)+ m^2\right)
u_m=0,\label{TT-tensor}
\end{align}
\end{subequations}
where $m$ is a 4D mass parameter and $k$ is the comoving
wavenumber along the 4D hypersurface.

Let us first consider Eq.~(\ref{TT-scalar}). If we write
$\alpha_m= \varphi_m/a(\eta)$ and use conformal time $\eta=- \int
(dt/a)$, then the wave equation on the brane reads as
\begin{equation}
\frac{d^2\varphi_m}{d\eta^2}+ \left[ -\frac{1}{a} \frac{d^2
a}{d\eta^2}+ a^2 m^2+ k^2\right]\varphi_m=0.
\end{equation}
With $a\propto e^{ t H}$(and hence $\eta= - 1/(a H) $), we get
\begin{equation}
\frac{d^2\varphi_m}{d\eta^2}+ \left[-\frac{2}{\eta^2}+
\frac{m^2}{\eta^2 H^2}+ k^2 \right]\varphi_m=0.
\end{equation}
The general solution is
\begin{equation}\label{time-evo-sol}
 \varphi_m(\eta; \vec{k})=\sqrt{\eta k}\, Z(\lambda,
\eta k), \quad  \nu \equiv \sqrt{\frac{9}{4}-\frac{m^2}{H^2}},
\end{equation}
where $Z(\nu, \eta k)$ is a linear combination of Bessel functions
of order $\nu$. One recovers the RS solution in the limit
$a(\eta)\to {\rm const} \equiv 1$, in which case $\varphi_m =
\exp(\pm i\omega t)$, with $\omega^2=k^2+m^2$. The perturbations
are over-damped for all light modes with $0< m< 3H/2$, while all
heavy modes with $m> 3H/2 $ oscillate and decay more rapidly and
the modes with $m^2<0$ do not exist. In a 4D de Sitter space, the
eigen modes satisfying $m^2 > 0$ are not localized on an inflating
(de Sitter) brane, see below.

\begin{figure}[!ht]
\centerline{\includegraphics[width=2.8in,height=1.6in]
{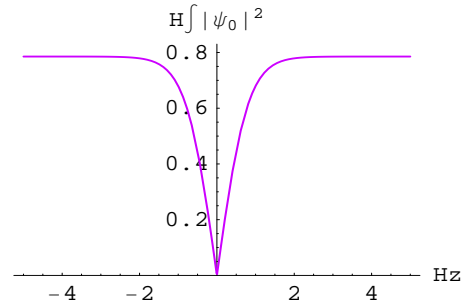}} \caption{ (color online). The plot of the function $H\int
|\psi_0|^2$ with $b\Z{0}=1$. } \label{zero-mode}
\end{figure}

\begin{figure}[!ht]
\centerline{\includegraphics[width=2.8in,height=1.6in]
{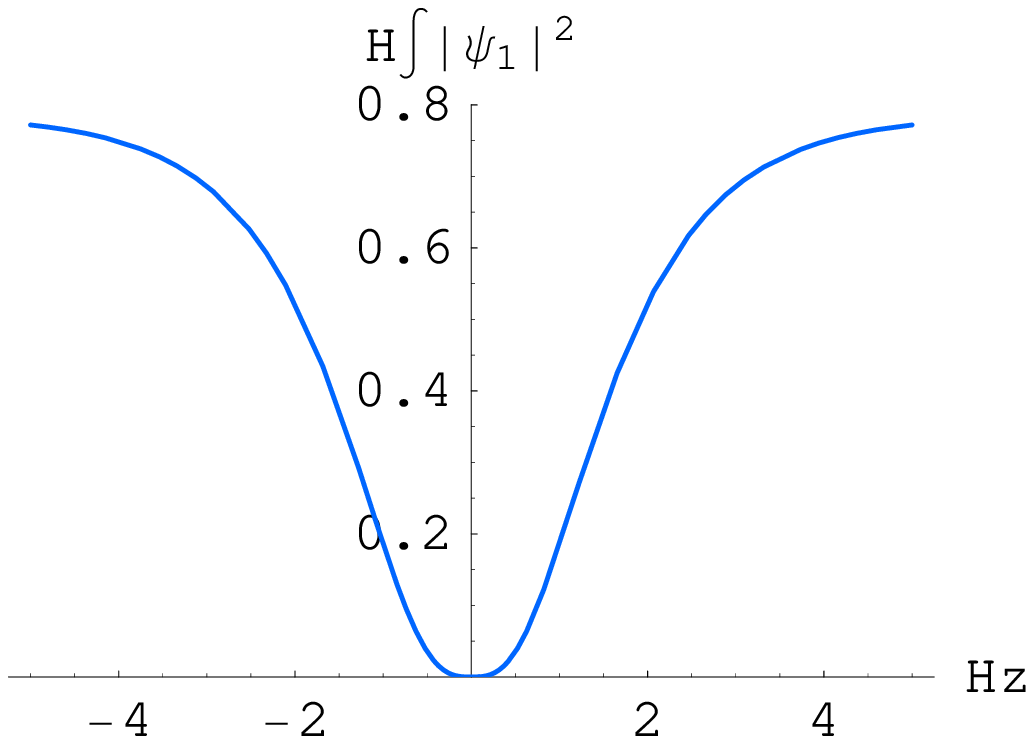}} \caption{(color online). The plot of the function $H\int
|\psi_1|^2$ with $b\Z{1}=1$.} \label{first-mode}
\end{figure}

Defining $u_m \equiv e^{-3A/2}\, \psi_m$, it is possible to
rewrite Eq.~(\ref{TT-tensor}) in a Schr\"odinger-like form
\begin{equation}
\frac{d^2 \psi_m}{d\phi^2} - V\, \psi_m  + m^2 \rho^2 \psi_m=0
\end{equation}
with the scalar potential
\begin{eqnarray}
V&\equiv & \frac{9}{4} {A'}^2+ \frac{3}{2}
A''-\frac{3H\rho\Phi_-}{\Phi_+} \delta(\phi-\phi_0)\nonumber \\
&=& \frac{9 \rho^2 H^2}{4} - \frac{15\lambda \rho^2
H^2}{\Phi_+^2}-\frac{6H\rho\Phi_-}{\Phi_+}
\delta(\phi-\phi_0),\label{full-poten2}
\end{eqnarray}
where we used the solution (\ref{sol-bg1}) and $\Phi_{\pm}=
e^{\rho H\phi} \pm \lambda e^{-\rho H \phi}$. Introducing a new
coordinate variable
$$
z\equiv \rho \phi - \ln \sqrt{\lambda},$$ we get
\begin{equation}
\frac{d^2\psi_m}{dz^2} - V \psi_m = - m^2 \psi_m,
\label{Schro-main}
\end{equation}
where
\begin{equation}
V = \frac{9H^2}{4}- \frac{15 H^2}{4 \cosh^2 (Hz)}-6 H
\tanh(Hz)\delta(z-z_c).\label{PositiveV}
\end{equation}
The brane is now located at $z=z_c$. The zero-mode solution
($m^2=0$) is given by
\begin{equation}
\psi\Z{0}(z)=\frac{b\Z{0}}{(\cosh(H z))^{3/2}},\label{sol1}
\end{equation}
which is clearly normalizable since
$$ \int_{-\infty}^{\infty} |\psi_0(z)|^2 dz= \frac{\pi b\Z{0}^2}{2H}.$$
There is one more bound state solution, i.e.,
\begin{equation}
\psi\Z{1}(z) = b\Z{1}\, \frac{\sqrt{\cosh^2(H z)-1}}{(\cosh(H
z))^{3/2}},\label{sol2}
\end{equation}
which is obtained by taking $m^2=2H^2$. This solution is also
normalizable. However, only the zero-mode solution ($m^2=0$) is
localized on the de Sitter brane. This can be seen by substituting
(\ref{sol1}) into Eq.~(\ref{Schro-main}) and comparing the
delta-function terms (cf Figs.~\ref{zero-mode}-\ref{first-mode}).


We also note that
\begin{subequations}
\begin{align}
\int \frac{|\psi\Z{0}|^2}{b\Z{0}^2} dz=
\frac{\tan^{-1}\tanh\frac{|Hz|}{2}}{H}+
\frac{{\rm sech}(Hz) \tanh \frac{|Hz|}{2}}{2H},\\
\int \frac{|\psi\Z{1}|^2}{b\Z{1}^2} dz=
\frac{\tan^{-1}\tanh\frac{|Hz|}{2}}{H}- \frac{{\rm sech}(Hz) \tanh
\frac{|Hz|}{2}}{2H},
\end{align}
\end{subequations}
The zero-mode graviton is localized on the brane. The first
excited state with mass $m^2=2H^2$ is normalizable but this mode
is not localized on the brane.

The general solution to Eq.~(\ref{Schro-main}) is given by
\begin{eqnarray}
&& \psi(z)= c\Z{1}\, X^{5/2} \, {}_2F_1 \left(\frac{5+2i\mu}{4},
\frac{5-2i\mu}{4}; \frac{1}{2}; 1-X^2\right)
\nonumber \\
&{}& \quad + \, c\Z{2}\, \sqrt{X^2-1}\, X^{5/2}\, \nonumber \\
&{}& \qquad \, {}_2F_1 \left(\frac{7+2i\mu}{4}, \frac{7-2i\mu}{4};
\frac{3}{2}; 1-X^2\right),
\end{eqnarray}
where $X =\cosh(Hz)$ and $\mu\equiv
\sqrt{\frac{m^2}{H^2}-\frac{9}{4}}=\pm i\nu$. The allowed values
of $m$ are quantized in units of $H$ or the index $\gamma\equiv
m^2/H^2$. Around the brane's position at $z\equiv z_c $,
satisfying $H z \ll 1$, the solution looks like
\begin{equation}
\psi(z) = c\Z{1} P_\mu + c\Z{2} Q_\mu,
\end{equation}
where
\begin{subequations}
\begin{align}
P_\mu = 1- \frac{3+2\gamma}{4}(H z)^2+ \frac{39+12\gamma+
4\gamma^2}{96} (H z)^4 + \cdots,\\
Q_\mu = H z -\frac{3+2\gamma}{12} (H z)^3 +
\frac{99+12\gamma+4\gamma^2}{480} (H z)^5 + \cdots.
\end{align}
\end{subequations}
Note that the condition $H z\ll 1$ signifies a cosmological scale
for which $H^{-1} \gg z$, i.e. the Hubble radius is much larger
than the radial extension of the fifth dimension.

\begin{figure}[!ht]
\centerline{\includegraphics[width=2.8in,height=2.0in] {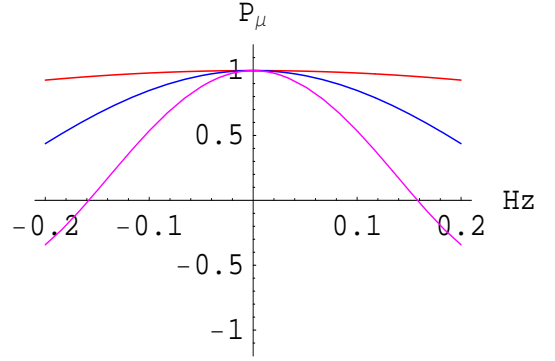}}
\caption{(color online). The plot of the function $P_\mu(z)$ with $\gamma=9/4, 30,
100$ (top to bottom). } \label{P-mu}
\end{figure}
\begin{figure}[!ht]
\centerline{\includegraphics[width=2.8in,height=2.0in] {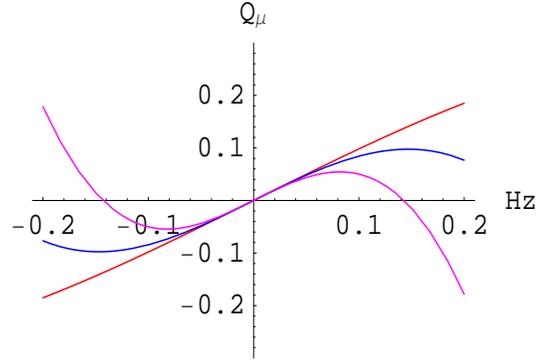}}
\caption{The plot of the function $Q_\mu(z)$ with $\gamma=9/4, 30,
100$ (top to bottom, in the range $Hz>0$).} \label{Q-mu}
\end{figure}

Further, using the following property:
\begin{eqnarray}
{}_2 F_1 (a,b; c; z) &=& (1-z)^{-b}\,{}_2F_1 (c-a, b; c;
\frac{z}{z-1}),\nonumber
\end{eqnarray}
the eigenfunctions for the massive continuous modes with $m^2\ge
2H^2$ can be expressed as
\begin{eqnarray}
\psi_m (z) &=& c\Z{1}\, \,X^{i\mu} \nonumber\\
&{}& \times \,{}_2F_1\left(-\frac{3+2i\mu}{4}, \frac{5-2i\mu}{4};
\frac{1}{2}, \frac{X^2-1}{X^2}\right)\nonumber \\
&{}& + \,c\Z{2}\,\frac{\sqrt{X^2-1}}{X} \,X^{i\mu} \nonumber
\\
&{}& \times \,{}_2F_1\left(-\frac{1+2i\mu}{4}, \frac{7-2i\mu}{4};
\frac{3}{2}, \frac{X^2-1}{X^2}\right),\nonumber\\
\label{main-soln}
\end{eqnarray}
where $X=\cosh(Hz)$. This result may be brought to the form given
in~\cite{Brevik-etal} [cf. Eq. (49)], applying Euler's
hypergeometric transformations. A direct comparison with the
result in~\cite{Kehagias02}, i.e., Eq.~(14), is less
straightforward as the result there was presented using Gaussian
normal coordinates. In the large $z H$ limit, we get
\begin{equation}
\psi\Z{Hz\to \infty} =c_1\, e^{i\mu z H} + c_2\, e^{-i\mu z H},
\end{equation}
where $\mu\equiv \sqrt{\frac{m^2}{H^2}-\frac{9}{4}}$. With
$c\Z{1}=0$, all heavy modes with $\mu> 0$ become oscillating plane
waves, which represent the de-localized KK modes (cf.
Figures.~\ref{P-mu}-\ref{Q-mu}). The time-evolution of the mode
functions of these heavy modes [cf. Eq.~(\ref{time-evo-sol})] shows
that they remain underdamped at late times $|\eta|\to 0$.

A couple of remarks are in order. 
In the
$\Lambda_5<0$ case, though the analysis closely follows the one
given above, the opposite sign of $\Lambda_5$ modifies some of the
results. For example, when $\Lambda_5<0$, the scalar potential $V$
in (\ref{Schro-main}) is given by
\begin{equation}
V = \frac{9H^2}{4}+ \frac{15 H^2}{4 \sinh^2 (Hz)}-6 H
\coth(Hz)\delta(z-z_c).
\end{equation}
The zero-mode solution ($m^2=0$) is given by
\begin{equation}
\psi\Z{0}(z)=\frac{c_0}{(\sinh(H z))^{3/2}}, \label{sol3}
\end{equation}
which is normalizable only when one allows a cutoff scale
satisfying $0< z_c \le z$ or $z\le z_c <0$.

The main finding in~\cite{Brevik-etal} was that the choices
$\Lambda_5>0$ and $\Lambda_5<0$ both may lead to a nomalizable
zero-mode graviton wave function, provided that one satisfies
\begin{eqnarray}
 \Lambda_5 \ge - \frac{\tau^2}{6 M_{(5)}^6} \quad  {\rm and} \quad
 \Lambda_5 \le \frac{\tau^2}{4 M_{(5)}^6}, \nn
\end{eqnarray} respectively, in the AdS$_5$ and dS$_5$ cases,
where $\tau$ is the 3-brane tension.
It is easy to understand the latter case, which follows from the
classical solution [cf. Eq.~(\ref{flat-brane-T})] plus the
positivity of the off-brane potential [cf . Eq.~(\ref{PositiveV})].
However, in the former case ($\Lambda_5<0$), it was also necessary
to have $Z_2$ symmetry, otherwise the zero-mode solution is
non-normalizable. Such constraints become much weaker (or even do
not exist), especially in dimensions, $D\ge 7$. Similar remarks
apply to mass gap or the mass of Kaluza-Klein states in higher
dimensions, which are determined in terms of a free parameter $M$
(associated with the size of fifth dimension) rather than in terms
of the 4D Hubble expansion parameter $H$.

\subsection{Linearized bulk equation}

Next we consider the bulk perturbations of the 5D metric, which
can be analyzed by considering a wave equation for a master
variable $\Omega \equiv \Omega(x^\mu; \phi)$ introduced
in~\cite{Koyama:04a}. In a 5D de Sitter background defined by
(\ref{5d-main-ac}) we find that
\begin{equation}\label{master-eq}
\frac{1}{\rho^2} \left(\frac{e^{-3A}}{a^3}\Omega'\right)^\prime
-\left(\frac{e^{-3A}}{a^3}\,\dot{\Omega}\right)^{\cdot} +
\left(\frac{\vec{\nabla}^2}{a^2}-\frac{e^{2A}}{\ell^2}\right)\frac{e^{-3A}}{a^3}\Omega=0.
\end{equation}
By using the following change of variable
\begin{equation}\label{new-def}
\Omega\equiv a(t)^3 e^{3A(\phi)}\, \tilde{\Omega},
\end{equation}
Eq.~(\ref{master-eq}) can be written as
\begin{eqnarray}
&&\frac{1}{\rho^2} \left(\tilde{\Omega}''+3A^\prime
\tilde{\Omega}'+3A'' \tilde{\Omega}\right)
-\left(\ddot{\tilde{\Omega}}+3\frac{\dot{a}}{a}\dot{\tilde{\Omega}}\right)
\nonumber \\
&&~~~~~~~~~~~~ \qquad +
\left(\frac{\vec{\nabla}^2}{a^2}-\frac{e^{2A}}{\ell^2}\right)\tilde{\Omega}=0.
\end{eqnarray}
By separating the newly defined master variable as
$\tilde{\Omega}(x^\mu,\phi) \equiv \sum \alpha_m(t)
u_m(\phi)\,e^{i k\cdot x}$, we get
\begin{subequations}
\begin{align}
\ddot{\alpha}_m + 3 \frac{\dot{a}}{a} \dot{\alpha}_m + \left(
\frac{k^2}{a^2} + m^2 \right) \alpha_m=0,\label{time-compo}\\
\frac{d^2 u_m}{d\phi^2} + 3 A^\prime u_m^\prime+ \left(3A''+ m^2
\rho^2 -\frac{\rho^2}{\ell^2}\, e^{2A}\right)
u_m=0.\label{wave-eq-phi}
\end{align}
\end{subequations}
The first equation above is the same as (\ref{TT-scalar}), so we
only have to consider the second equation.

Defining $u_m \equiv e^{-3A/2}\, \Phi_m$, we get
\begin{equation}
 \frac{d^2 \Phi_m}{d\phi^2} - V\, \Phi_m +  m^2 \rho^2 \Phi_m=0,
\end{equation}
with the off-brane potential $V$ of the form
\begin{eqnarray}
U&\equiv & \frac{9}{4} {A'}^2- \frac{3}{2} A'' +
\frac{\rho^2}{\ell^2} \, e^{2A}\nonumber \\
&=&\frac{9 \rho^2 H^2}{4} + \frac{\lambda \rho^2
H^2}{\Phi_+^2},\label{full-poten}
\end{eqnarray}
where we used the solution (\ref{sol-bg1}) and $\Phi_{\pm}=
e^{\rho H\phi} \pm \lambda e^{-\rho H \phi}$. Defining $z\equiv
\rho \phi -\ln\sqrt{\lambda}$, as before, we get
\begin{equation}\label{bulk-wave1}
- \frac{d^2 \Phi_m}{dz^2} + V\, \Phi_m =  m^2 \Phi_m,
\end{equation}
with the scalar potential $V$ of the form
\begin{eqnarray}
V =\frac{9 H^2}{4} + \frac{H^2}{4\cosh^2(Hz)}.
\end{eqnarray}
This equation can also be obtained directly from
Eq.~(\ref{master-eq}) but using $\Omega\equiv \sum  \alpha_m(t)
u_m(\phi)$ and $u_m\equiv e^{-3A/2} \Phi_m$, see,
e.g.,~\cite{Koyama:04a}. The main difference, as compared to the
result in AdS$_5$ spacetimes, is the sign of the second term
above. The general solution to (\ref{bulk-wave1}) is given by
\begin{eqnarray}
&& \Phi_m(z)= c_1\, \left(\cosh(Hz)\right)^{1/2} \nonumber
\\
&{}& \qquad \times \, {}_2F_1 \left(\frac{1+2\nu}{4},
\frac{1-2\nu}{4}; \frac{1}{2}; -\sinh^2(H z)\right)
\nonumber \\
&{}& \quad + \, c_2\, |\sinh(Hz)|\left(\cosh(Hz)\right)^{1/2}\nonumber\\
&{}& \qquad \times \, {}_2F_1 \left(\frac{3+2\nu}{4},
\frac{3-2\nu}{4}; \frac{3}{2}; -\sinh^2(H z)\right),
\end{eqnarray}
where $\nu\equiv \sqrt{\frac{9}{4}-\frac{m^2}{H^2}}$. Again, there
are 2 bound state solutions: (i) $\nu=3/2$ ($m^2=0$) and
\begin{eqnarray}
&& \Phi_0(z) = c_1 \sqrt{X}\sqrt{X^2-1}
~~~~~~~~~~~~~~~~~~\nonumber \\
&& \quad +\, c_2 \sqrt{X}\left(1-\sqrt{X^2-1} \tan^{-1}
\frac{1}{\sqrt{X^2-1}}\right),\label{eigen-zero}
\end{eqnarray}
(ii) $\nu=1/2$ ($m^2=2 H^2$) and
\begin{equation}
\Phi\Z{1}(z)= \sqrt{X} \left( c_1 +c_2 \tan^{-1}
\frac{1}{\sqrt{X^2-1}}\right),\label{eigen-first}
\end{equation}
where $X\equiv \cosh(Hz)$. Both these solutions are
non-normalizable. This result is desirable as it implies that a
massless bulk scalar mode may not be localized on a de Sitter
brane. Note that there are no any tachyonic or growing modes
localized on the brane either.

\subsection{Projected Weyl tensor}

The nonexistence of arbitrarily light KK excitations can be seen
also by considering the wave equation for a projected 5D Weyl
tensor, which is given by~\cite{Sasaki2000a}
\begin{equation}\label{5d-pertb}
\left[e^{A}\frac{\partial}{\partial z} e^{-A}
\frac{\partial}{\partial z} + \Box_4 - 4
H^2\right]\hat{E}_{\mu\nu}=0,
\end{equation}
where $\hat{E}_{\mu\nu} \equiv e^{2A}\, E_{\mu\nu}$, ${E}_{\mu\nu}
\equiv C^A\,_{\mu B \nu} n^A n^B$ is the projected 5D Weyl tensor,
$n^A$ is the vector unit normal to the brane, $\Box_4 \equiv
g^{\mu\nu} D_\mu D_\nu$ is the 4-dimensional d'Alembertian with
respect to the metric $g_{\mu\nu}$, and $D_\mu$ is the covariant
derivative. With a separation of variable $\hat{E}_{\mu\nu}=
\Psi(\phi) Y_{\mu\nu}^{(m)} (x^\mu)$, Eq.~(\ref{5d-pertb}) yields
\begin{subequations}
\begin{align}
 \left[  e^{A} \frac{\partial}{\partial z} e^{-A}
\frac{\partial}{\partial z} - 2 H^2 \right] \Psi_m = - m^2
\Psi_m,\label{main-Schro1}\\
\left[ \Box_4 - m^2  - 2 H^2 \right] Y_{\mu\nu}^{(m)}=0,
\end{align}
\end{subequations}
where $m$ is a 4D mass parameter, which has been introduced here
as a separation constant. In this formalism, we clearly see that
there is a mode $\Psi={\rm constant}$ with $m^2 = 2H^2$ that
trivially satisfies the equation.

Defining $\Psi_m \equiv e^{A/2}  \Phi_m$, we can rewrite the
off-brane wave equation~(\ref{main-Schro1}) as
\begin{equation}\label{Schrodinger2}
- \frac{d^2\Phi_m}{dz^2} + V\, \Phi_m =  m^2  \Phi_m,
\end{equation}
where
\begin{eqnarray}
V &\equiv & \frac{{A'}^2}{4}- \frac{A''}{2} + 2 H^2 = \frac{9
H^2}{4} + \frac{H^2}{4\cosh^2(Hz)}.
\end{eqnarray}
This is the same potential as in (\ref{bulk-wave1}). The mode
$m^2=2H^2$ translates to $\Phi_m\propto e^{-A/2}\propto
\cosh(Hz)$, and it is the first eigenmode and is obtained from
(\ref{eigen-first}) with $c_2=0$.

\subsection{Correction to Newton's law}

To estimate the correction to Newton's force law generated by a
discrete tower of Kaluza-Klein modes, one may go to the thin brane
limit, i.e. $H^{-1} \to 0$, but keeping the ratio $z_c/H^{-1}$
finite. One also assumes that the matter fields in the
4-dimensional theory is smeared over the width of the brane and
the brane thickness is smaller compared with the bulk curvature,
$H^{-1} < \ell$, so $H\ell > 1$. Under this approximation, the
gravitational potential between 2 point-like sources of masses
$M_1$ and $M_2$ located on the brane is modified via exchange of
gravitons living in 5 dimensions as
\begin{eqnarray}
U(r)&=&G_4 \frac{M_1 M_2}{r} + M_{(5)}^{-3}\int_{m}^{\infty} dm\,
\frac{M_1
M_2\, e^{-m r}}{r} \,|\Phi_m (z_c)|^2 \nonumber \\
&\simeq & \frac{G_4 M_1 M_2}{r} \left[1+ \frac{ H^2 \ell^3}{\alpha
r} \sum_i  e^{-m_i r} \left(1+ {\cal O} \left(\frac{1}{H r}\right)
\right) \right],\nonumber \\
\end{eqnarray}
where $m_i \ge \sqrt{2} H$, $r$ is the distance between the two
pointlike sources, and $\alpha$ is a constant of order unity. This
result qualitatively agrees with that given
in~\cite{Aguilar10a,Ghoroku}. In~\cite{Kehagias02} a concern was
raised, especially, in the dS$_5$ case, that the correction to the
gravitational potential due to the massive KK states may dominate
for $r < H^{-1}< \ell $, leading to a 5 dimensional behavior which is
not Newtonian. But one should also note that the corrections to
the gravitational potential are suppressed by a factor of $\sum_i
e^{-m_i r}$. As a result, there is perhaps no restriction in
taking $\ell \gg r$, provided that the KK modes are sufficiently
heavy. For instance, with $m_i\gtrsim {\rm TeV}\sim 10^{-15}~{\rm
cm}$, $\alpha\sim {\cal O}(1)$, the correction term may not show
up unless we probe a sufficiently small distance scale, like
$r\sim 10^{-12}~{\rm cm}$.

In the presence of matter or gauge fields, one may require a more
complete analysis, involving the effects of the overlap of the
gravitational modes with the matter modes, but we will not
consider this case here.

\section{Generalization to higher dimensions}

For a consistent description of gravity plus gauge field theories,
one may require models with more than one extra extra dimensions,
and the world around us could have up to $n=6$ (or $n=7$) extra
spatial dimensions, if our universe is described by string theory
(or M-theory). In the following, as some canonical examples, we
will consider the $n=2$ and $n=3$ cases, but it is straightforward
to generalize the present discussion to $D=10$ or $D=11$
dimensions.

First, we allow 2 extra dimensions ($n=2$) and write the 6D
action as
\begin{equation}\label{6d-gravi}
S_{\rm grav}  = M_{(6)}^4 \int d^6{x} \sqrt{-g}\,R,
\end{equation}
where $M_{(6)}$ is the 6D Planck mass. It is not difficult to
check that the metric ansatz
\begin{equation}\label{dd-main-ac}
ds_{6}^2 = \frac{1}{K^{2p}} \left(-dt^2+a(t)^2 d\vec{x}_{3,k}^2 +
\frac{p^2 {K^\prime}^2}{H^2 K^2} dz^2 + L^2 K^{2p}
d\theta^2\right)
\end{equation}
where $K^\prime = dK/dz$ and $0\le \theta \le 2\pi$, solves the 6D
Einstein equations with the 4D scale factor
\begin{equation}
a(t)=\frac{c\Z{0}}{2}\, e^{H t}+ \frac{k}{2c\Z{0}H^2}\,e^{-Ht}.
\end{equation}
In the above, $K(z)$ is an arbitrary function of $z$.

As a simple example, we take $K(z)\equiv \cosh(M z)$, where $M$
has a mass dimension of one. Then, under the dimensional
reduction, from $D=6$ to $D=4$, we get
\begin{equation}
M\Z{(6)}^4 \int d^6{x} \sqrt{-g} R= M_{\rm Pl}^2 \int d^4 x
\sqrt{-\hat{g}_4} \left(R_{(4)}- \Lambda\Z{4}\right),
\end{equation}
where $\Lambda_4=6 H^2$ and
\begin{equation}
M_{\rm Pl}^2= M\Z{(6)}^4 \frac{p L}{H} \int_0^{2\pi} d\theta
\int_{-\infty}^{\infty}\frac{K^\prime\,dz}{K^{3p+1}}= \frac{4\pi L
M\Z{(6)}^4 }{3H}.
\end{equation}
Especially, in $D=6$, if we introduce a bulk cosmological term,
then we will also require another source term, e.g., 2-form gauge
field, in order have an explicit solution. In dimensions $D\ge 7$,
however, a bulk cosmological constant can be introduced into the
Einstein action without considering other source terms.

Next we note that the following 7-dimensional metric,
\begin{equation}
ds_{7}^2 =\frac{1}{F(z)}\left( -dt^2 + a(t)^2 d\vec{x}_{3,k}^2 +
G(z) dz^2 + E d\Omega_{2}^2 \right),
\end{equation}
solves the Einstein field equations following from
\begin{equation}
S= M\Z{(7)}^5 \int d^7{x}\sqrt{-g}
\left(R-2\Lambda\Z{b}\right),\label{7DwithCC}
\end{equation}
when
\begin{equation}
G(z)= \frac{15{F^\prime(z)}^2}{36H^2 F(z)^2-4\Lambda_b F(z)},
\quad E=\frac{1}{3H^2},\label{sol-G}
\end{equation}
where $F^\prime = dF(z)/dz$. There can exist a large class of 4D
de Sitter solutions with different choices of $F(z)$.
Generalization of this solution to higher dimensions is
straightforward.

Here we consider 2 physically interesting examples.

\subsection*{Example 1}
Take
\begin{equation}
F(z)\equiv F\Z{0}^2 \cosh^2 ( M z ).
\end{equation}
The dimensionally reduced action reads as
\begin{eqnarray}
&& M\Z{(7)}^5 \int d^7{x} \sqrt{-g} (R-2\Lambda\Z{b})\nonumber \\
&& \qquad = M_{\rm Pl}^2 \int d^4 x \sqrt{-\hat{g}_4}
\left(R_{(4)}- \Lambda\Z{4}\right),
\end{eqnarray}
where $\Lambda\Z{4}=6 H^2$,
\begin{equation}
M_{\rm Pl}^2 = M\Z{(7)}^5 \frac{2\pi M \sqrt{15}}{9 F\Z{0}^5  H^3}
I(z) \end{equation} and
\begin{equation}
I(z)\equiv \int_{-\infty}^{\infty}\frac{\sqrt{\cosh^2(M z)-1}\,
dz}{\cosh^5(M z)\sqrt{\cosh^2(M z)-\beta}},\label{3d-warped}
\end{equation}
with
$$ \beta \equiv \frac{\Lambda\Z{b}}{9 H^2 F\Z{0}^2}. $$
With $\Lambda\Z{b}=0$, so $\beta=0$, one can easily evaluate the
above integral and find that $I(z)=2/(5M)$. In the
$\Lambda\Z{b}>0$ case, we require $0< \beta \le 1$. With $\beta
=1$, we get $I(z)=3\pi/(8M)$. That is, provided that $0<\beta<1$,
the integral converges and its value ranges between $2/(5M)$ and
$3\pi /(8M)$.

In case the Hubble expansion parameter $H$ is large, or
equivalently, when $|\beta| < 1$, the mass reduction formula is
well approximated by
\begin{equation} M_{\rm Pl}^2 \sim \frac{M_{(7)}^5}{H^3 F_0^5}.
\end{equation}
However, when $H$ becomes small, or one considers a large
cosmological distance scale, then one should allow a negative bulk
cosmological term. The integral converges for any negative value
of $\beta$. In the limit $|\beta|\gg 1$, the mass reduction
formula is well approximated by the relationship
\begin{equation} M_{\rm Pl}^2 \sim \frac{M_{(7)}^5}{H^2
F\Z{0}^4}\, \frac{1}{(-\Lambda\Z{b})^{1/2} }.
\end{equation}
This result may be analyzed further by taking $H \sim
10^{-60}~M_{\rm Pl}$, especially, if one wants to tune $\Lambda_4$
to the present value of 4D cosmological constant. To satisfy
phenomenological constraints, such as, $M_{(7)}\gtrsim {\rm TeV}$
and $\left(-\Lambda\Z{b}\right)^{1/2}\lesssim M_{(7)}$, one then
has to allow $F\Z{0}$ to take a reasonably large value, $F\Z{0}
\gtrsim 10^{14} \gg 1$. A similar constraint could arise in five
dimensions as well; see, for example,~\cite{Ghoroku}, where  a
realistic construction required $ H^2/\Lambda_b \sim 10^{-15}$ (in
the $D=7$ case, the ratio $H^2/\Lambda_b$ equals to $1/9 F_0$). A
constraint like this becomes much weaker when one applies the
model to explain the early universe inflation. For instance, with
$H \sim 10^{-5} ~M_{\rm Pl}$, we get $M_{\rm Pl}\sim 10^{-3}
M_{{\rm Pl} (7)}/F\Z{0} $, in which case $F\Z{0}$ may be taken to
be small, say $F\Z{0}\sim {\cal O}(1)$.

To say anything further in a concrete way, we need to have some
physical information about the constant $F\Z{0}$, which may be
related to the D-dimensional dilaton coupling.

\subsection*{Example 2}
Take
\begin{equation}
F(z)\equiv F\Z{0}^2\, \exp\left(2 \arctan \left(e^{M z}\right)
\right).
\end{equation}
The warp factor $F(z)$ and the function $G(z)$ are regular
everywhere. The integral (\ref{3d-warped}) is now modified as
\begin{equation}
I(z)\equiv \int \frac{\left(1-\beta(z)\right)^{-1/2}\, dz}{4
\cosh(M z)\,\exp\left(5\arctan \left(e^{Mz}\right)\right)},
\end{equation}
where
\begin{equation}
 \beta(z) \equiv \frac{\Lambda\Z{b}}{9 H^2 F\Z{0}^2}\,
\exp\left(-2\arctan\left(e^{Mz}\right)\right).
\end{equation}
Since $\arctan(x) \to \pm \frac{\pi}{2}$ as $x\to \pm \infty$, the
above integral gives a finite result provided that $\Lambda\Z{b} <
9 H^2 F_0^2\,e^\pi $, leading to a dynamical mechanism of
compactfication. The choice $\Lambda\Z{b}<0$ is fairly safe from
the viewpoint of metric regularity, or the smoothness of $G(z)$.

The classical solutions given above are regular everywhere,
$-\infty< z<\infty $. However, at the linearized level, one may be
required to introduce a cutoff along $z$-direction. In the
$\Lambda_b=0$ case, the off-brane Schr\"odinger-type potential and
the zero-mode solution ($m^2=0$) are given, respectively, by
\begin{equation} V(z)= \frac{21}{16} \left(\frac{F'}{F}\right)^2 +
\frac{3}{4} \left(\frac{F''}{F'}\right)^2 - \frac{1}{2}
\frac{F'''}{F'}
\end{equation}
and
\begin{equation}
\psi_0(z) \propto \left(
\frac{dF(z)}{dz}\right)^{-1/2}\,F(z)^{-3/4}.
\end{equation}
For instance, with $F(z)\propto \cosh^2 (M z)$, we obtain
\begin{equation}
\psi_0(z)= \frac{c\Z0}{\cosh^2 (Mz) |\sinh (M z)|^{1/2}}.
\end{equation}
Here one is required to take $0< z_c \le z$ or $z \le z_c < 0$ so
as to ensure normalizability of the zero-mode wavefuction. For a
choice like $F(z)\propto \exp\left(\alpha \arctan \exp(Mz)\right)$
(where $\alpha$ is arbitrary), however, there arises no such
restriction.

In the $\Lambda_b \ne 0$ case, the off-brane wave equation for
tensor perturbations satisfies
\begin{equation}
\frac{d^2 u_m}{dz^2} - \left(\frac{5F'}{2F} + \frac{G'}{2G}
\right) \frac{du_m}{dz} + m^2 u_m(z)=0.\label{7d-perturb}
\end{equation}
$G(z)$ was defined in (\ref{sol-G}). By defining
\begin{equation}
u\Z{m}(z) \propto \left(\frac{F' F^2}{\sqrt{9H^2
F-\Lambda_b}}\right)^{1/2} \psi(z),
\end{equation}
we can bring Eq.~(\ref{7d-perturb}) into the standard Schrodinger
equation; the explicit expression of the potential $V$, which is
rather lengthy, is not important for our discussion here. We only
mention that the zero-mode solution is given by
\begin{equation}
\psi_0(z) \propto \frac{\left(9 H^2 F(z)-
\Lambda_b\right)^{1/4}}{F(z) \sqrt{dF/dz}}.
\end{equation}
As for classical solutions, there exists a constraint that
$\Lambda_b< 9 H^2 F(0)$. As another rather simple example, we may
take $F(z) \propto e^{Mz}$ and restrict the radial coordinate in
the range $0\le  z < \infty$. Indeed, the zero-mode graviton
wave function is normalizable with all of the above choices of
$F(z)$.

All the results above can easily be generalised to higher
dimensions, including the 10- and 11-dimensional models inspired by
string/M-theory~\cite{Ish09,Ish10b}. We refer to~\cite{Ish10d} for
further discussions on localization of gravity. The method can
also be extended to a class of thick domain walls (or de Sitter
brane solutions) in gravity coupled to a bulk scalar
field~\cite{DeWolfe99,Kanti-etal,Maeda:2000,Kobayashi}. In the
next section, we only consider the $D=5$ case. The discussions in
general D dimensions will be given elsewhere.

\section{A thick domain-wall solution in D=5}

The 5D gravitational action is modified as
\begin{equation} S_5=\frac{1}{2} \int d^5{x} \sqrt{-g}
\left[\frac{R}{\kappa_5^2} - g^{AB}
\partial_A \phi \partial_B \phi - 2 V(\phi) \right],
\end{equation}
where $\kappa_5^2 \equiv 1/M_{(5)}^3$. The full 5D warped metric
is
\begin{equation} ds^2 = e^{2A(z)}
\left( ds_4^2 + dz^2\right).
\end{equation}
As above we search for a class of 4D FLRW cosmologies with de
Sitter expansion, characterized by the metric
\begin{equation}
ds_4^2 = L^2 \Big[ -dt^2 + a(t)^2 \left(\frac{dr^2}{1-k  r^2} +
r^2\,d\Omega_2^2\right)\Big],
\end{equation}
where $L$ is a scale associated with the size of the physical 3+1
spacetime. For simplicity, we begin with the case $k=0$ and
$a(t)\propto e^{Ht}$. This simplification is also justified in a
realistic cosmological scenario, at least, for late time
cosmologies.

Next, consider that $\phi$ depends only on the extra dimension,
$\phi\equiv \phi(z)$. The 5D field equations then take the form
\begin{eqnarray}
{\phi^\prime}^2 &=& \frac{3}{\kappa_5^2}
\left({A^\prime}^2-A^{\prime\prime}-\frac{H^2}{L^2}\right),\label{eq1}\\
V(\phi) &=& \frac{3 \,e^{-2A}}{2 \kappa_5^2}
\left(\frac{3 H^2}{L^2}-3 {A^\prime}^2 - A^{\prime\prime}\right),\label{eq2}\\
\frac{dV}{d\phi} &=& e^{-2A} \left(\phi^{\prime\prime}+3A^\prime
\phi^\prime \right),\label{eq3}
\end{eqnarray}
where ${}^\prime \equiv d/dz$. These equations admit a series of
solutions with different choice of $\phi(z)$ or $V(\phi(z))$. As
an illustrative case, we look for the standard domain-wall type
solution:
\begin{equation}
\kappa_5 \phi= \phi\Z0 - \phi\Z{1} \arcsin \tanh \left( \frac{
Hz}{L \delta}\right),\label{scalar-anz}
\end{equation}
where $\phi_i$ and $\delta$ are dimensionless constants. The
solution for the warp factor is given by
\begin{equation}
A(z)= A\Z{0}- \delta \ln \cosh \left(\frac{H}{L} \frac{
z}{\delta}\right).
\end{equation}
The coefficient $\phi\Z{1}$ in (\ref{scalar-anz}) is fixed as
$\phi\Z{1}= \sqrt{3\delta(1-\delta)}$. Note that $\phi$ behaves as
a canonical scalar field only when $0<\delta<1$. This will remain
our choice in the following discussion. The scalar potential is
\begin{equation}
V(\phi)=\frac{3(1+3\delta) H^2}{2\delta \kappa_5^2 L^2}
\left[\cos^2\left(\frac{\phi\Z{0}-\kappa_5
\phi}{\phi\Z{1}}\right)\right]^{(1-\delta)}.
\end{equation}
The case $\delta=1$ is special for which $\phi$ and hence
$V(\phi)$ are constants. This potential has a maximum at $z=0 $.
This is seen by noting that
\begin{equation}
\cos^2 \left(\frac{\phi\Z0-\kappa\phi}{\phi\Z1}\right)= {\rm
sech^2}\left(\frac{H}{L} \frac{ z}{\delta}\right),
\end{equation}
where we used the solution (\ref{scalar-anz}). The height of the
potential decreases with the expansion of the universe since $L$
is required to be larger ($H$ becomes smaller) at late epochs.
These solutions are available with a nonzero 3D curvature, i.e.
with the scale factor $a(t)=(a\Z0/2) \,e^{Ht} + (k/2 H^2 a\Z{0})\,
e^{-Ht}$.

\begin{figure}[!ht]
\centerline{\includegraphics[width=3.2in,height=2.0in]
{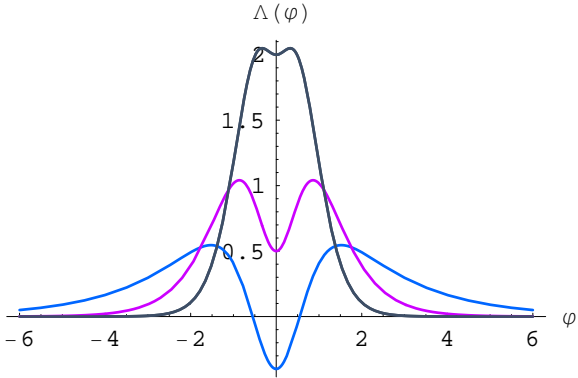}} \caption{(color online). The plot of the function
$\Lambda(\varphi)$ with $\delta=1$, $0.5$, and $0.2$ (top to
bottom). In the $\delta=0$ case, the potential is unbounded,
leading to a diverging $\Lambda\Z{4}$.} \label{4d-poten}
\end{figure}

To find an effective 4D potential, we shall consider a
dimensionally reduced action. From the solution given above, we
derive
\begin{eqnarray}
S_{\rm eff} &= & \int e^{3A(z)} dz \int d^4{x}\sqrt{-g_4}\nn \\
&{}& \times \left(\frac{R_4}{2\kappa_5^2} - \frac{\left(6
{A^\prime}^2 + 4A^{\prime\prime}\right)}{\kappa_5^2}-\frac{1}{2}
{\phi^\prime}^2 - V(\phi) e^{2A}\right)\nonumber
\\
&=& \frac{1}{\kappa_5^2} \int e^{3A(z)} dz \int d^4{x}
\sqrt{-g_4}\nn\\
&{}& \qquad \times \left(\frac{R_4}{2} - (3{A^\prime}^2 +
A^{\prime\prime}) - \frac{3 H^2}{L^2}\right)\nonumber\\
&=& \frac{1}{2\kappa_4^2} \int d^4{x} \sqrt{-g_4} \, {R}_4 - \int
d^4{x} \sqrt{-g_4}\, \Lambda\Z{4},\nonumber \\
\end{eqnarray}
where
\begin{eqnarray}
\frac{1}{\kappa_4^2} &\equiv & \frac{e^{3A\Z0}}{\kappa_5^2}
\int_{-\infty}^{\infty} \frac{dz}{\left[ \cosh\left(\frac{H}{L}
\frac{ z}{\delta}\right)\right]^{3\delta}}\nn \\
&=& \frac{e^{3A\Z0}\,L\delta }{\kappa_5^2 H}
\int_{\varphi_1}^{\varphi_2} \frac{d\varphi}{\left(
\cosh{\varphi}\right)^{3\delta}},
\end{eqnarray}
and
\begin{eqnarray}
\Lambda\Z{4} &\equiv & \frac{e^{3A\Z0}}{\kappa_5^2}\frac{H}{L}
\int_{\varphi\Z{1}}^{\varphi\Z{2}}
 \frac{6\delta \cosh^2 \varphi-3\delta
 -1}{\left(\cosh{\varphi}\right)^{2+3\delta}}\,d\varphi \nn \\
&=& \frac{e^{3A\Z0}}{\kappa_5^2} \frac{H}{L} \int
\Lambda(\varphi)\,d\varphi,
\end{eqnarray}
where we introduced a new variable $\varphi\equiv \frac{ H
z}{L\delta} $. For $0<\delta<1$ the potential is Mexican-hat type
(cf. Figure\ref{4d-poten}).

Note that the integral
\begin{equation}
I\Z{\delta}\equiv \int_{-\infty}^{\infty}
\frac{d\varphi}{\left(\cosh{\varphi}\right)^{3\delta}}=\frac{\sqrt{\pi}\,
\Gamma\left[3\delta/2\right]}{\Gamma\left[(3\delta+1)/2\right]}
\end{equation}
is finite as long as $\delta>0$. For instance
$I\Z{\delta=0.5}=2.3958$ and $I\Z{\delta=1}=\pi/2$. This all shows
that
\begin{equation}
M_{\rm Pl}^2 \sim  \frac{L M_{(5)}^3\, e^{3A\Z0}}{H} = {\cal R}
M_{(5)}^3,
\end{equation}
where ${\cal R}\equiv (L/H)\,e^{3A\Z{0}}$ is the effective size of
the fifth dimension. In our notation above, $H^{-1}$ measures the
Hubble radius, but $L$ is a dimensionless constant. As is clear,
to get a small ${\cal R}$ we shall take an exponentially small
$e^{A\Z{0}}$. For $\delta>0$ the effective 4D potential is
positive. We rule out the case $\delta=0$ because in this case the
4D Planck mass is not finite. Moreover, $\int
\Lambda(\varphi)=-2$, which implies that $\Lambda\Z{4}$ is
negative in 4-dimensions.

We have shown that, as in the model without a scalar Lagrangian,
the effective 4D Newton's constant is finite despite having a
noncompact direction.

\medskip

\section{Conclusion}

Finally, we summarize the main results in the paper.

Simple five-dimensional brane-world models defined in an AdS$_5$
spacetime have been known to provide a rich phenomenology for
exploring some of the intriguing ideas that are emerging from
string/M-theory, such as, AdS/CFT correspondence, AdS holography
and mass hierarchies.

The replacement of AdS$_5$ spacetime by dS$_5$ spacetime, along
with replacement of a flat 3-brane by a physical 4D universe, is
found to give us new problems and new possibilities. The problem
is that the embedding of dS$_4$ in dS$_5$ may be viewed as a
$O(4)$ symmetric bubble described by a Coleman-De Luccia
instanton~\cite{Coleman1980}, in which case the size of the bubble
may not exceed the radius of dS$_5$. As a result, a constraint
like $ c H^{-1}\le \ell_{\rm dS}$ could bring the 5D Planck mass
down to TeV scale or even much lower. This problem can easily be
overcome by introducing two or more extra dimensions, along with a
higher-dimensional bulk cosmological term and background fluxes.

Another issue could be that dS$_5$ allows a foliation by a flat
space but that is a spacelike hypersurface. There is no way to cut
dS$_5$ by Minkowski spacetime. That is, in a cosmological setting,
a flat 4D Minkowski spacetime is not a solution to 5D Einstein
equations, if the bulk spacetime is de Sitter. This is in contrast
to the results in Randall-Sundrum brane-world models in AdS$_5$
spacetimes. But this is anyway not a real problem since the
Universe has probably never gone through a phase of being close to
a static universe or a flat 3-brane.

There is perhaps no necessity of having a Minkowski spacetime
embedded in a dS$_5$ spacetime as long as the massless graviton
wave function becomes normalizable on a 4D de Sitter spacetime,
which is indeed the case within the model considered in this
paper. We have shown that the effective 4-dimensional Planck
mass derived from the fundamental D-dimensional Planck mass can be
finite because of the large but finite warped volume and also the
large world-volume of a de Sitter brane (i.e., the physical
Universe), implying that a $Z_2$ symmetry available to 5D
brane-world models in an AdS background can be simply relaxed.

We have followed the most general approach for obtaining
cosmologies with de Sitter expansion from the higher-dimensional
Einstein equations, which could yield characteristic linear
4-dimensional spacetime sizes of many orders of magnitude bigger
than linear sizes in extra coordinates. We have obtained some
interesting classical gravity solutions that compactify
higher-dimensional spacetime ($D\ge 5$) to produce a
Robertson-Walker universe with de Sitter type expansion plus 1
extra noncompact direction. We have also shown that such models
can admit both an effective 4-dimensional Newton constant that
remains finite and a normalizable graviton wave function.

Compared to some known results in five dimensions, for instance,
in Refs.~\cite{Kanti-etal,Brevik-etal,Kehagias02,Kobayashi}, we
find that both the classical and linearized solutions are less
constrained in higher spacetime dimensions. De Sitter brane-world
models in higher dimensions appear to be less restrictive (more
viable) also on a phenomenological ground. For instance, in
dimensions $D\ge 7$, the mass gap or the mass of KK modes may be
determined in terms of a free parameter $M$ (associated with the
size of fifth dimension) rather than in terms of the 4D Hubble
expansion parameter $H$. The latter is the situation in five
dimensions, where the lightest KK mode will have mass $m=\sqrt{2}
H$, which is not sufficiently massive in low energy or in the
present Universe with $H\sim 10^{-60}\,M_{\rm Pl}$. Moreover, in
spacetime dimensions $D\ge 7$, the classical solutions (with or
without a bulk cosmological constant) could lead to a spontaneous
compactification of the extra dimensions.

\medskip

{\sl Acknowledgements:} It is a pleasure to thank Kazuya Koyama,
Kei-ichi Maeda, M. Sami, Takahiro Tanaka, and Roy Maartens for
insightful discussions and correspondences on topics related to
the theme of this paper. This work is supported by the Marsden
fund of the Royal Society of New Zealand.


\end{document}